\documentstyle[twoside,fleqn,espcrc2,epsfig]{article}

\newcommand{\br}{\mbox{Br}\,}

\newcommand{\kmm}{K_L\rightarrow\mu^+\mu^-}
\newcommand{\kee}{K_L\rightarrow e^+ e^-}
\newcommand{\emm}{\eta\rightarrow\mu^+\mu^-}
\newcommand{\epmm}{\eta'\rightarrow\mu^+\mu^-}
\newcommand{\eee}{\eta\rightarrow e^+ e^-}
\newcommand{\epee}{\eta'\rightarrow e^+ e^-}
\newcommand{\pee}{\pi^0\rightarrow e^+ e^-}

\newcommand{\mesgg}{P\rightarrow\gamma\gamma}
\newcommand{\mesll}{P\rightarrow l^+l^-}
\newcommand{\epll}{\eta'\rightarrow l^+l^-}

\newcommand{\kll}{K_L\rightarrow l^+l^-}
\newcommand{\kgg}{K_L\rightarrow\gamma\gamma}

\newcommand{\gev}{\mbox{GeV}}
\newcommand{\rms}{\rm\scriptsize}

\title{
\vspace{-2cm}
{\small\rightline{FTUV/98-83}
\rightline{IFIC/98-84}}
\vspace{1.2cm}
Short-- and long--distance contributions to the rare decay
$K_L\rightarrow\mu^+\mu^-$} 
\author{D.\ G\'omez Dumm and A.\ Pich \\ \hfill \\
\noindent 
Departament de F\'{\i}sica Te\`orica, IFIC,
Universitat de Val\`encia -- CSIC,\\
Dr.\ Moliner 50, E-46100 Burjassot, Val\`encia, Spain\thanks{Talk given
at the Euroconference on Quantum Chromodynamics (QCD 98), Montpellier,
France, 2-8 Jul 1998.}}

\begin{document}

\thispagestyle{empty}

\begin{abstract}
The interplay between short-- and long--distance contributions to the
$\kmm$ decay amplitude is analyzed. The long--distance piece is
estimated using chiral perturbation theory techniques and large--$N_C$
considerations, leading to a consistent description of the $\pee$,
$\emm$ and $\kmm$ branching ratios \cite{GP:98}.
\end{abstract}

\maketitle


\section{INTRODUCTION}

The rare decay $\kmm$ has deserved a significant theoretical interest
during the last three decades. It represents a
potentially important channel to study the weak interaction within
the Standard Model (SM), as well as possible effects of new physics,
mainly in connection with flavour--changing neutral currents and CP
violation.

This decay proceeds through two distinct mechanisms: a long--distance
contribution from the $2\gamma$ intermediate state
and a short--distance part, which in the SM arises from one--loop
diagrams ({\it $W$ boxes, $Z$ penguins}) involving the weak gauge bosons.
Since the short--distance amplitude is sensitive to the presence of a 
virtual top quark, it could be used to improve our present knowledge
on the quark--mixing factor $V_{td}$; moreover, it offers a
window into new--physics phenomena. 

The short--distance SM amplitude is well--known \cite{GL:74}.
Including QCD corrections at the next-to-leading logarithm order,
it implies \cite{BF:97}:
\begin{eqnarray}
\label{eq:sd}
  \br(\kmm)_{\mbox{\rms SD}}\! & =\! & 0.9 \times 10^{-9} \;
  (\rho_0-\bar\rho)^2 \; \nonumber \\
  & & \hspace{-1.7cm} \times
  \left({\overline{m}_t(m_t)\over 170\;\gev}\right)^{3.1}\;
  \left({\left| V_{cb}\right|\over 0.040}\right)^4 ,
\end{eqnarray}
where $\rho_0 \approx 1.2$ and $\bar\rho\equiv\rho\, (1-\lambda^2/2)$,
with $\rho$ and $\lambda$ the usual quark--mixing parameters, 
in the Wolfenstein parametrization.
The deviation of $\rho_0$ from 1 is due to the charm contribution.
Using the presently allowed ranges for $m_t$ and the quark--mixing
factors, one gets \cite{BF:97}
\mbox{$\br(\kmm)_{\rms SD} = (1.2\pm 0.6) \times 10^{-9}$}.
If this number is compared with the measured rate \cite{Ak:95}
\begin{equation}
\label{eq:exp}
  \br(\kmm ) = (7.2\pm 0.5) \times 10^{-9} \, ,
\end{equation}
it is seen that the decay process is strongly dominated by the
long--distance amplitude.

Clearly, in order to extract useful information about the short--distance 
dynamics it is first necessary to have an accurate (and reliable)
determination of the
$K_L\rightarrow\gamma^\ast\gamma^\ast\rightarrow\mu^+\mu^-$
contribution.

Let us consider the normalized ratios
\begin{eqnarray}
\label{eq:R_def}
  R(\mesll ) \! & \equiv \! & {\br(\mesll)\over\br(\mesgg)} \nonumber \\
  & & \hspace{-1.5cm} = \, 2\beta\,
  \left({\alpha\over\pi}{m_l\over M_P}\right)^2\;
  \left| F(\mesll )\right|^2 \, ,
\end{eqnarray}
where
$\beta\equiv\sqrt{1-4m_l^2/M_P^2}$.
The on--shell $2\gamma$ intermediate state generates
the absorptive contribution \cite{MRS:70}
\begin{equation}
\label{eq:ImR}
	\mbox{Im}\ [F(\mesll )] = {\pi\over 2\beta}\, 
	\ln{\left({1-\beta\over 1+\beta}\right)} \, ,
\end{equation}
which, taking into account the measured \mbox{$\kgg$} branching ratio, leads to
the so-called {\it unitarity bound}:
\begin{eqnarray}
\label{eq:abs}
  \br (\kmm ) \! & \geq \! & \br (\kmm )_{\mbox{\rms Abs}} \nonumber \\
  & = \! & (7.07\pm 0.18)\times 10^{-9} \, .
\end{eqnarray}
Comparing this result with the experimental value in Eq.~(\ref{eq:exp}),
we see that $\br (\kmm )$ is almost saturated by this absorptive
piece. Then, one immediate question is whether
the small room left for the dispersive contribution,
$\br (\kmm )_{\mbox{\rms Dis}} = (0.1\pm 0.5)\times 10^{-9}$,
can be understood dynamically.

\section{$\kgg$}

The obvious theoretical framework to perform a well-defined analysis of
the long--distance $\kmm$ amplitude is chiral perturbation theory (ChPT).
Unfortunately, the chiral symmetry constraints are not powerful enough to
make an accurate determination of the dispersive
part~\cite{EP:91,DIP:97,VA:97}.

The problem can be easily understood by looking at the $\kgg$ amplitude,
\begin{equation}
\label{eq:Kgg}
 A(\kgg ) = c(q_1^2,q_2^2)\; \varepsilon^{\mu\nu\rho\sigma}\,
 \epsilon_{1\mu} \epsilon_{2\nu} q_{1\rho} q_{2\sigma} \, ,
\end{equation}
which, at lowest order in momenta, proceeds through the chain
$K_L\to \pi^0,\eta,\eta'\to 2\gamma$.
The lowest--order ---$O(p^4)$--- 
chiral prediction can only generate a constant
form factor $c(q_1^2,q_2^2)$, corresponding to the decay into
on-shell photons \cite{ENP:92}:
\begin{equation}
\label{eq:c_Kgg}
 c(0,0) = {2 G_8 \alpha f_\pi  \over\pi}\,
 \left(c_\pi+c_\eta+c_{\eta'}\right) \, ,
\end{equation}
where $c_{\pi,\eta,\eta'}$ stand for the $\pi$, $\eta$ and $\eta'$
pole contributions respectively, and the global parameter
$G_8\equiv 2^{-1/2} G_F V^{\phantom{*}}_{ud} V_{us}^*\, g_8$
characterizes \cite{EPR:88} the strength of the weak $\Delta S=1$
transition.

In the standard $SU(3)_L\otimes SU(3)_R$ ChPT, the $\eta'$ contribution
is absent, and $(c_\pi+c_\eta) \propto (3 M_\eta^2 + M_\pi^2 - 4 M_K^2)$,
which vanishes owing to the Gell-Mann--Okubo mass relation.
The physical $\kgg$ amplitude is then a higher--order
---$O(p^6)$--- effect in the chiral counting, which makes difficult to
perform a reliable calculation.

The situation is very different if one considers the
large--$N_C$ limit, in which the symmetry of the
effective theory is enlarged to $U(3)_L\otimes U(3)_R$ \cite{HLPT:97}, 
and the singlet $\eta_1$ field is included. The large mass of the
$\eta'$ originates in the $U(1)_A$
anomaly which, although formally of $O(1/N_C)$, is numerically
important.
Thus, it makes sense to perform a combined chiral expansion \cite{LE:96}
in powers
of momenta and $1/N_C$, around the nonet--symmetry limit,
but keeping the anomaly contribution (i.e.\ the $\eta'$ mass) together
with the lowest--order term.
In fact, the usual successful description of the $\eta/\eta'\to 2\gamma$
decays corresponds to the lowest--order contribution within this
framework, plus some amount of symmetry breaking through
$f_\eta\not= f_{\eta'}\not=f_\pi$.
The mixing between the $\eta_8$ and $\eta_1$ states
provides a large enhancement of
the  $\eta\to 2\gamma$ amplitude, which is clearly
needed to understand the data.

Although the resulting numerical prediction for $c(0,0)$ contains
several theoretical uncertainties
(values of $f_\eta$ and $f_{\eta'}$, deviations from the nonet symmetry
limit, accuracy in $\theta_P$ and $|G_8|$), it is seen that the
actual value of the $\kgg$ rate can be easily fitted
within the $U(3)_L\otimes U(3)_R$ framework for a reasonable choice
of the parameters. The amplitude is found to be dominated by the pion pole,
owing to a destructive interference between the $c_\eta$ and $c_{\eta'}$
contributions.

\section{DISPERSIVE $\kll$ AMPLITUDE}

The description of the $\kgg$ transition with off-shell photons is
a priori more complicated because the $q_{1,2}^2$ dependence of the	
form factor originates from higher--order terms in the chiral lagrangian.
This is the reason why only model--dependent estimates of the
dispersive $\kll$ transition amplitude have been obtained so far.
At lowest--order in momenta, \ $c(q_1^2,q_2^2) = c(0,0)\, $; thus, the 
(divergent) photon
loop can be explicitly calculated up to a global normalization, which
is determined by the known absorptive piece (i.e. by the
experimental value of $c(0,0)$). The model--dependence appears in
the local contributions from direct $K_L l^+ l^-$ terms
in the chiral lagrangian 
\cite{EP:91,VA:97} (allowed by symmetry considerations),
which reabsorb the loop divergence.

It would be useful to have a reliable determination in some
symmetry limit. The large--$N_C$ description of
$K_L\rightarrow\gamma^\ast\gamma^\ast$ provides such
a possibility \cite{GP:98}. At leading order, this process occurs through
the $\pi^0,\eta,\eta'$ poles, as represented in Fig.\ 1. Therefore, the
problematic electromagnetic loop in Fig.\ 1(a) is actually the same
governing the decays $\pee$ and $\emm$, and the unknown local contribution
(Fig.\ 1(b)) can be fixed from the measured rates for these transitions
\cite{SLW:92,ABM:93}. It can be seen \cite{SLW:92} that the same
combination of local chiral couplings shows up in both decays, leading to
a relation that is well satisfied by the data. Moreover, this combination
is also the relevant one for the $\epll$ transition in the large $N_C$
limit \cite{GP:98}.

\begin{figure}[htb]
\begin{center}
\vspace{-.7cm}
\epsfig{file=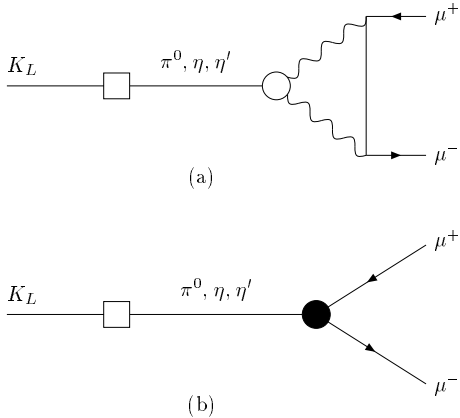}
\end{center}
\vspace{-.7cm}
\caption{(a) Photon loop and (b) associated counterterm contributions
to the $\kmm$ process.}
\vspace{-.4cm}
\end{figure}

Nonet symmetry should provide a good estimate of the ratio
$R(\kll)$. Since $\kgg$ is dominated by the pion pole, we can
expect that symmetry--breaking corrections would play a rather small
role. In this limit, the dispersive amplitude for all $R(\mesll)$ is
given~by
\begin{eqnarray}
\label{eq:disp}
\mbox{Re\ }[F(\mesll )] \! & = \! & 
{1\over 4\beta} \ln^2{\left({1-\beta\over 1+\beta}\right)} \nonumber \\
& & \hspace{-3.7cm} + {1\over \beta} Li_2\left({\beta-1\over\beta +1}\right)
+ {\pi^2\over 12\beta} + 3 \ln{\left({m_l\over\mu }\right)} 
+ \chi(\mu)
\, ,
\end{eqnarray}
where $\chi(\mu)$ is the relevant local contribution, renormalized
in the $\overline{\mbox{MS}}$ scheme.
The $\mu$ dependence of the $\chi(\mu)$ and $\ln{\left(m_l/\mu\right)}$
terms compensate each other, so that the total amplitude
is $\mu$--independent.

\section{RESULTS}

Table \ref{tab:chi} shows the fitted values of $\chi (M_\rho)$ from
the three measured ratios $R(\pee )$, \mbox{$R(\emm )$} and $R(\kmm )$.
Subtracting the known absorptive contribution, the experimental data 
provide two possible solutions for each ratio;
they correspond to a total positive (solution 1) or negative (solution 2)
dispersive amplitude. We see from the Table that the second solution from
the decay $\pee$ is clearly ruled out; owing to the smallness of the
electron mass,
the logarithmic loop contribution dominates the dispersive amplitude,
which has then a definite positive sign
(an unnaturally large and negative value of $\chi(M_\rho)$ is needed
to make it negative).
The large experimental errors do not allow to discard at this point
any of the other solutions: the remaining value from $\pee$ is consistent
with the results from the $\emm$ and $\kmm$ decays, and these are
also in agreement with each other if the same solution (either the first
or the second) is taken for both. We see that, in any case, the three
experimental ratios are well described by a common value of
$\chi (M_\rho)$. In this way, the experimentally observed small
dispersive contribution to the $\kmm$ decay rate fits perfectly well
within the large--$N_C$ description of this process.

\begin{table}[tb]
\caption{Fitted values of $\chi (M_\rho)$ from different $R(\mesll)$ ratios.
The numbers quoted for $\kmm $ refer to the difference
$\chi (M_\rho) - \delta\chi_{\mbox{\protect\rms SD}}$.}
\label{tab:chi}
\begin{tabular*}{75mm}{@{}l@{\extracolsep{\fill}}rr}
\hline
& [Solution 1] & [Solution 2] \\ \hline
$\pee $ & $4\,{}^{+4}_{-6}$ & $-24\pm 5$ \\
$\emm $ & $5.5\,{}^{+0.8}_{-1.0}$ & $-0.8\,{}^{+1.0}_{-0.8}$  \\
$\kmm $ & $3.3\,{}^{+0.9}_{-0.7}$ & $1.9\,{}^{+0.7}_{-0.9}$ \\
\hline
\end{tabular*}
\vspace{-.2cm}
\end{table}

We have not considered up to now the short--distance
contribution to the $\kmm$ decay amplitude \cite{BF:97}. This can be done
through a shift of the effective
$\chi(M_\rho)$ value\footnote{The relative sign between the short-- and
long--distance dispersive amplitudes is fixed by the known positive sign
of $g_8$ in the large--$N_C$ limit \cite{PR:96}.}:
\begin{equation}
\chi(M_\rho)_{\rms eff} = \chi(M_\rho) - \delta\chi_{\rms SD}
\, ,
\end{equation}
$$
\delta\chi_{\mbox{\rms SD}} \approx 1.7\, \left(\rho_0 -\bar\rho\right)\,
\left({\overline{m}_t(m_t)\over 170\;\gev}\right)^{1.56}\,
  \left({\left| V_{cb}\right|\over 0.040}\right)^2 .
$$ 
For the allowed range $|\bar\rho|\leq 0.3$, one has
$\delta\chi_{\mbox{\rms SD}}\approx 1.8\pm 0.6$, which allows to
exclude the solution 2 for $\chi(M_\rho)$ obtained from $\emm$.
The solution 1, on the contrary, is found to be compatible with
the results from $\kmm$, and can be used to get a constraint for
$\delta\chi_{\mbox{\rms SD}}$. Indeed, taking as the best determination
\begin{equation}
\chi(M_\rho) = 5.5\, {}^{+0.8}_{-1.0}\; ,
\label{eq:chi}
\end{equation}
the first solution for $\kmm$ leads to
\begin{equation}
\delta\chi_{\mbox{\rms SD}} =
2.2\, {}^{+1.1}_{-1.3} \; ,
\label{eq:dchi}
\end{equation}
in agreement with the $\delta\chi_{\mbox{\rms SD}}$ value quoted above.
The second solution for $\kmm$ appears to be less favoured, yielding
$\delta\chi_{\mbox{\rms SD}}=3.6\pm 1.2$; this shows a discrepancy of
about $1.4\,\sigma$ with the short--distance estimate. Notice that the
precision of the result in (\ref{eq:dchi}) is still relatively low. However,
the errors could be reduced by improving the measurements of the $\emm$
and $\kmm$ branching ratios.

Once the local contribution to the $\mesll$ decay amplitude has
been fixed, it is possible to obtain definite predictions for the
decays into $e^+ e^-$ pairs:
\begin{equation}
\begin{array}{ll}
\mbox{Br}(\pee) = (8.3 \pm 0.4)\times 10^{-8} \, , \\
\mbox{Br}(\eee) = (5.8 \pm 0.2)\times 10^{-9} \, , \\
\mbox{Br}(\kee) = (9.0 \pm 0.4)\times 10^{-12} \, .
\end{array}
\end{equation}
The predicted $\kee$ decay rate has been confirmed by the recent
BNL-E871 measurement~\cite{BNL:98}, $\mbox{Br}(\kee)= (8.7^{+5.7}_{-4.1})
\times 10^{-12}$.

In the same way, the amplitudes corresponding to the $\eta'$ decays are
found to be $\mbox{Br}(\epee) = (1.5 \pm 0.1)\times 10^{-10}$ and
$\mbox{Br}(\epmm) = (2.1 \pm 0.3)\times 10^{-7}$. However, in view of the
large mass of the $\eta'$, these predictions could receive important
corrections from higher--order terms in the chiral lagrangian.

\hfill

To summarize, we have shown that in the nonet symmetry limit it is
possible to make a reliable determination of the ratios $R(\mesll)$,
at lowest non-trivial order in the chiral expansion. A consistent
picture of all measured $\mesll$ modes is obtained within the SM.
In the case of the $\kmm$ decay, the present data allow to get a
constraint for the short--distance amplitude, which could be improved
by more precise measurements of the $\emm$ and $\kmm$ branching ratios.
Although a more detailed investigation of the underlying theoretical
uncertainties is still required, this analysis offers a new possibility
for testing the flavour-mixing structure of the Standard Model.

This work has been supported in part by CICYT and DGESIC, Spain
(Grants No.\ AEN-96-1718 and PB97-1261) and by the European Union
TMR programme (Contracts No.\ ERBFMRX-CT98-0169 and ERBFMBI-CT96-1548).


\begin{thebibliography}{30}

\bibitem{GP:98} D.\ G\'omez Dumm and A.\ Pich, Phys.\ Rev.\ Lett.\ {\bf 80},
4633 (1998).

\bibitem{GL:74} M.\ K.\ Gaillard and B.\ W.\ Lee, Phys.\ Rev.\ D
{\bf 10}, 897 (1974).

\bibitem{BF:97} A.\ J.\ Buras and R.\ Fleischer, in Heavy Flavors II,
eds.\ A.\ J.\ Buras and M.\ Linder, Advanced Series on Directions in
High Energy Physics -- Vol.\ 15 (World Scientific, Singapore, 1998),
p.\ 65, {\tt hep-ph/9704376}.

%
\bibitem{Ak:95} T.\ Akagi {\em et al.}, Phys.\ Rev.\ D {\bf 51}, 2061 (1995);
  A.\ P.\ Heinson {\em et al.}, Phys.\ Rev.\ D {\bf 51}, 985 (1995).

\bibitem{MRS:70} B.\ R.\ Martin, E.\ de Rafael and J.\ Smith, Phys.\ Rev.\
D {\bf 2}, 179 (1970).

\bibitem{EP:91} G.\ Ecker and A.\ Pich, Nucl.\ Phys.\ {\bf B366}, 189 (1991).

\bibitem{DIP:97} G.\ D'Ambrosio, G.\ Isidori and J.\ Portol\'es,
Phys.\ Lett.\ {\bf B423}, 385 (1998).

\bibitem{VA:97} G.\ Valencia, Nucl.\ Phys.\ {\bf B517}, 339 (1998).

\bibitem{ENP:92} G.\ Ecker, H.\ Neufeld and A.\ Pich, Phys.\ Lett.\
  {\bf B278}, 337 (1992).

\bibitem{EPR:88} G.\ Ecker, A.\ Pich and E.\ de Rafael,
  Phys.\ Lett.\  {\bf B189}, 363 (1987);
  Nucl.\ Phys.\ {\bf B291}, 692 (1987),
  {\bf B303}, 665 (1988).

\bibitem{HLPT:97} P.\ Herrera-Sikl\'ody {\em et al.}, Nucl.\ Phys.\
 {\bf B497}, 345 (1997); Phys.\ Lett.\ {\bf B419}, 326 (1998).

\bibitem{LE:96} H.\ Leutwyler, Phys.\ Lett.\ {\bf B374}, 163 (1996).

\bibitem{SLW:92} M.\ J.\ Savage, M.\ Luke and M.\ B.\ Wise, Phys.\
   Lett.\ {\bf B291}, 481 (1992).

\bibitem{ABM:93} Ll.\ Ametller, A.\ Bram\'on and E.\ Mass\'o, Phys.\ Rev.\
   D {\bf 48}, 3388 (1993).

\bibitem{PR:96} A.\ Pich and E.\ de Rafael, Phys.\ Lett.\
 {\bf B374}, 186 (1996).

\bibitem{BNL:98} D.\ Ambrose {\em et al.}, {\tt hep-ex/9810007}.

\end{thebibliography}
\end{document}